\documentclass[twocolumn,amsmath,amssymb,aps,pra,floatfix]{revtex4}
\usepackage{bm}
\usepackage{amsfonts}
\usepackage{amsmath}
\usepackage{graphicx}
\usepackage{epstopdf}
\begin{document} 
\title{Geometrical optics and geodesics in thin layers}
\author{Tomasz Rado\.zycki}
\email{t.radozycki@uksw.edu.pl}
\affiliation{Faculty of Mathematics and Natural Sciences, College of Sciences, Cardinal Stefan Wyszy\'nski University, W\'oycickiego 1/3, 01-938 Warsaw, Poland} 
\begin{abstract}
The propagation of a light ray in thin layer (film) within geometrical optics is considered. It is assumed that the ray is captured inside the layer due to reflecting walls or total internal reflection (in the case of a dielectric layer). It has been found that for a very thin film (the length scale is imposed by the curvature of the surface at a given point) the equations describing the trajectory of the light beam are reduced to the equation of a geodesic on the limiting curved surface. There have also been found corrections to the trajectory equation resulting from the finite thickness of the film. Numerical calculations performed for a couple of exemplary curved layers (cone, sphere,  torus and catenoid) confirm that for thin layers the light ray which is repeatedly reflected, propagates along the curve close to the geodesic but as the layer thickness increases, these trajectories move away from each other. Because the trajectory equations are complicated non-linear differential equations, their solutions show some chaotic features. Small changes in the initial conditions result in remarkably different trajectories. These chaotic properties become less significant the thinner the layer under consideration.
\end{abstract}
\maketitle

\section{Introduction}

Physics played out in thin structures becomes increasingly important due to the miniaturization in electronics, the appearance of new materials like graphene and intensive development of nanophysics. In particular the issue of the propagation of electromagnetic waves trapped in thin films (layers) is expected to find various possible applications in optoelectronics, optical communication or integrated optics~\cite{tam1,hun,tam2}. This trapping can be achieved by reflecting walls of various shapes or in thin dielectric layers possessing some refractive index due to the phenomenon of total internal reflection as it happens in the so called open waveguides~\cite{sc,yeh,bal,carson,chew,collin,marcuse}.

Light propagation in such structures is often depicted within the two-dimensional reduced theory of electromagnetism~\cite{hillion,lapidus,zwiebach,trq} which is used as a model. It is true that the electromagnetic modes in a layer to some extent may be described in that way but certain limitations have been established which affect the propagation of electromagnetic waves~\cite{brb,trq}. 

In the present work we concentrate in turn on geometrical optics. A light ray traveling in the two-di\-men\-sional curved surface as a rule follows a geodesic due to the Fermat's principle. In a real layer, even very thin, this ray is continuously reflected between the walls, consequently following a very complicated path. The aim of this work is to clarify how these two trajectories are related to each other and under what circumstances the physical one may be treated as a geodesic drawn on the two-dimensional surface (i.e. one of the walls). 

In some special cases the wave equations inferred from Maxwell's electromagnetism in thin curved media have been derived by a limiting procedure~\cite{batz,lai}. As regards the geometrical optics the property of the rays traveling along geodesics has been exploited in the so called geodesic lenses, where the light propagates in a medium of space-dependent refractive index~\cite{bra,li}. However, the problem we tackle below up to our knowledge has not been addressed and it is commonly taken for granted that the propagation in a thin layer corresponds to the two-dimensional geodesic~\cite{ri,corn}. It should also be noted, that some experimental results involving the propagation of real light beams in such circumstances have been obtained, pointing at the role of the intrinsic curvature of a given medium~\cite{schul}. The extrinsic curvature turned out to be inessential as long as the light is trapped within layer.

The present paper is organized as follows. In Sec.~\ref{s} the geometrical problem of a light ray repeatedly reflected between two curved walls is considered. Assuming that the layer in between is thin, a differential equation for the trajectory as projected onto one of the walls is derived and the deviations from the corresponding geodesics are identified. As expected these deviations become more significant for thicker layers particularly in the regions where the curvature is large. 

The layer in question will be parametrized as ${\bm r}(x^1,x^2,w)$ where the first two parameters serve as Euler's coordinates on the surface and the derivative $\partial{\bm r}/\partial w$ points in the direction perpendicular to the walls (i.e. $w$ numerates different surfaces). By the appropriate choice of the scale for the parameter $w$ we can ensure that $|\partial{\bm r}/\partial w|=1$. 

The metric tensor referred to throughout the paper as $g_{ij}$ is defined {\em on the surface}. It is then the two-dimensional object ($i,j=1,2$) expressed through  $x^1$, $x^2$ and of course $w$  (which is however constant on a given surface) and not that in the $3D$ space. The dependence on $w$ means that $g_{ij}$'s in general differ on the distinct boundaries. In order to find the true trajectory, one has to abandon the surface at least infinitesimally, and therefore not surprisingly the derivative $\partial g_{ij}/\partial w$ appears in the formulas.

In this paper we deal with layers of uniform thickness or, more precisely, of the two limiting surfaces (walls) defined by the relations ${\bm r}(x^1,x^2,w)$ and ${\bm r}(x^1,x^2,w+\delta w)$, where $\delta w$ is small and constant while moving along surfaces. All layers considered as examples comply with these requirements. Surely one can imagine more complicated situations in which a given layer has different thickness at various points. Then the limiting procedure would be in a sense `local'. Such situations remain beyond the scope of the present work.

We also assume that the layers, although very thin, are thick with respect to the wavelengths of the propagating light. Otherwise the rules of the geometrical optics could not be applied. The second assumption, which seems reasonable too, is that the length scales imposed by surface normal curvatures (e.g. the curvature radii) are large with respect to the layer thickness. In other words one can say that the layer is relatively `smooth' without sharp warps. Without this assumption we could not expect the light ray to be trapped within the dielectric layer since it might happen that at a given point the angle of incidence could become inferior to the critical angle.

In Sec.~\ref{num} the several particular examples of truly curved surfaces, both with positive and negative intrinsic curvature, are dealt with in detail. They are: cone, sphere, torus and catenoid. The deviations from the geodesics on these surfaces are presented on the plots found by the numerical solutions of the trajectory equations. 

Throughout the paper the Einstein's summation convention over repeated indices is constantly used.

\section{The equation of the light-ray trajectory}
\label{s}

Let us assume that a light ray travels along the layer being reflected by the walls. The key equation for its trajectory is
\begin{eqnarray}
\hat{{\bm k}}'=&&\!\!\!\!\hat{{\bm k}} -2(\hat{{\bm k}}\!\cdot\! \partial_w {\bm r}(x'^1,x'^2, w+\delta w))\nonumber\\
&&\!\!\!\!\times\partial_w {\bm r}(x'^1,x'^2, w+\delta w),
\label{kpkq}
\end{eqnarray} 
where $\hat{{\bm k}}$ and $\hat{{\bm k}}'$ are unit vectors identifying the directions of the incident and reflected rays. The Eq.~(\ref{kpkq}) simply constitutes the statement that angle of incidence for a ray being reflected at ${\bm r}(x'^1,x'^2, w+\delta w)$ equals the angle of reflection. Upon multiplying both sides by $k'$, we can rewrite it in the form
\begin{eqnarray}
{\bm k}'=&&\!\!\!\!\lambda\big[{\bm k} -2({\bm k}\!\cdot\! \partial_w {\bm r}(x'^1,x'^2, w+\delta w))\nonumber\\
&&\!\!\!\!\times\partial_w {\bm r}(x'^1,x'^2, w+\delta w)\big],
\label{kpk}
\end{eqnarray} 
where $\lambda=k'/k$ is a certain normalization constant since $k'$ and $k$ can in general be of unequal lengths. Squaring both sides of this equation, it can be easily verified that
\begin{eqnarray}
{\bm k}'^2=&&\!\!\!\!\lambda^2\big[{\bm k} -2({\bm k}\!\cdot\! \partial_w {\bm r}(x'^1,x'^2, w+\delta w))\nonumber\\
&&\!\!\!\!\times\partial_w {\bm r}(x'^1,x'^2, w+\delta w)\big]^2=\lambda^2{\bm k}^2,
\label{sqk}
\end{eqnarray}
where the normalization~(\ref{norm}) of the normal vector has been used.

\begin{figure}[h]
\begin{center}
\includegraphics[width=0.48\textwidth,angle=0]{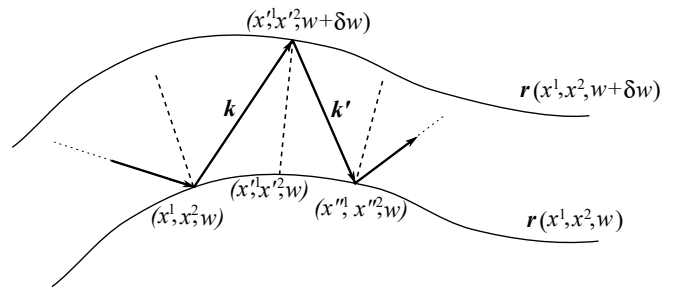}
\end{center}
\caption{The path of the light ray reflected by the walls of the layer defined by ${\bm r}(x^1,x^2,w)$ and ${\bm r}(x^1,x^2,w+\delta w)$.}
\label{refle}
\end{figure}

In order to shorten the notation, we henceforth  denote:
\begin{subequations}
\begin{align}
{\bm r}&={\bm r}(x^1,x^2,w),\label{defrp1}\\
{\bm r}'&={\bm r}(x'^1,x'^2,w),\label{defrp2}\\
{\bm r}''&={\bm r}(x''^1,x''^2,w),\label{defrp3}
\end{align}
\end{subequations}
and also 
\begin{equation}
\delta x^i=x'^i-x^i,\;\;\;\; \delta x'^i=x''^i-x'^i.\label{defde}
\end{equation}

The tangent vectors to the surface ${\bm r}(x^1,x^2,w)$ at a given point correspond to the derivatives with respect to $x^1$ and $x^2$ as to which the following notation will be used:
\begin{subequations}\label{conv}
\begin{align}
\frac{\partial}{\partial x^i}\, {\bm r}(x^1,x^2, w)&=\partial_i{\bm r}(x^1,x^2, w)=\partial_i{\bm r},\label{conv1}\\
\frac{\partial}{\partial x'^i}\,  {\bm r}(x'^1,x'^2, w)&=\partial_i'{\bm r}(x'^1,x'^2, w)=\partial_i'{\bm r'},\label{conv2}
\end{align}
\end{subequations}
and similarly for the higher derivatives. According to the chosen parametrization spoken of in the Introduction the normal vector
\begin{equation}
\partial_w{\bm r}(x^1,x^2, w)=\frac{\partial}{\partial w}\, {\bm r}(x^1,x^2, w),
\label{now}
\end{equation}
is normalized to unity 
\begin{equation}
\partial_w{\bm r}\!\cdot\! \partial_w{\bm r}=1,
\label{norm}
\end{equation}
and orthogonal to the tangent vectors:
\begin{equation}
\partial_i{\bm r}\!\cdot\! \partial_w{\bm r}=0,\;\; i=1,2.
\label{ortog}
\end{equation}

As to the higher derivatives with respect to $w$ one can show that
\begin{equation}
\partial_w^2{\bm r}\!\cdot\! \partial_w{\bm r}=\frac{1}{2}\, \partial_w (\partial_w{\bm r}\!\cdot\! \partial_w{\bm r})=0,\label{sede1}
\end{equation}
thanks to the normalization, and that
\begin{eqnarray}
\partial_w^2{\bm r}&&\!\!\!\!\!\cdot\! \partial_i{\bm r}=\partial_w (\partial_w{\bm r}\!\cdot\! \partial_i{\bm r})-\partial_w{\bm r}\!\cdot\! \partial_w\partial_i{\bm r}\label{sede2}\\
&&=-\partial_i(\partial_w{\bm r}\!\cdot\! \partial_w{\bm r})+\partial_i\partial_w{\bm r}\!\cdot\! \partial_w{\bm r},\nonumber
\end{eqnarray}
i.e.
\begin{eqnarray}
\partial_w^2{\bm r}\!\cdot\! \partial_i{\bm r}&&\!\!\!\! =\partial_i\partial_w{\bm r}\!\cdot\! \partial_w{\bm r}=\partial_w (\partial_i{\bm r}\!\cdot\! \partial_w{\bm r})-\partial_i{\bm r}\!\cdot\! \partial^2_w{\bm r}\nonumber\\
&&\!\!\!\!=-\partial_w^2{\bm r}\!\cdot\! \partial_i{\bm r}=0.
\label{sede3}
\end{eqnarray}
Since the vectors $\partial_w{\bm r}$ and $\partial_i{\bm r}$ ($i=1,2$) constitute a basis in the three-dimensional space and the vector $\partial_w^2{\bm r}$ turns out to have the null decomposition onto them, so one can infer that it is simply a null vector. Consequently also the higher derivatives with respect to $w$ vanish (for arbitrary $x^1,x^2$). Thereby, the parametrization of the surface ${\bm r}(x^1,x^2, w)$ has to linearly depend on $w$. It also means that 
\begin{equation}
\partial_w{\bm r}(x^1,x^2, w)=\partial_w{\bm r}(x^1,x^2, w+\delta w)
\label{norwd}
\end{equation}
which justifies the notation adopted in Fig.~\ref{refle}. The normal passing through the point ${\bm r}(x^1,x^2, w)$ is the same as that passing through ${\bm r}(x^1,x^2, w+\delta w)$.

Let us now project the vectorial equation~(\ref{kpk}) subsequently onto $\partial_w{\bm r'}$ and $\partial_i'{\bm r'}$, obtaining the following set
\begin{subequations}\label{sla}
\begin{align}
\partial_w{\bm r'}\!\cdot\! {\bm k}'&=-\lambda \partial_w{\bm r'}\!\cdot\! {\bm k},\label{slaw}\\
\partial_i'{\bm r'}\!\cdot\! {\bm k}'&=\lambda \partial_i'{\bm r'}\!\cdot\! {\bm k},\;\; i=1,2.\label{slai}
\end{align}
\end{subequations}
upon exploiting the normalization and orthogonality conditions (\ref{norm}) and~(\ref{ortog}) as well as~(\ref{norwd}).
Eliminating $\lambda$, one gets the fundamental equation for the light-ray trajectory:
\begin{equation}
(\partial_i'{\bm r'}\!\cdot\! {\bm k}')\, (\partial_w{\bm r'}\!\cdot\! {\bm k})+(\partial_i'{\bm r'}\!\cdot\! {\bm k})\, (\partial_w{\bm r'}\!\cdot\! {\bm k'})=0.
\label{ell}
\end{equation}

From Fig.~\ref{refle} it is obvious that
\begin{subequations}\label{kkp}
\begin{align}
{\bm k}&={\bm r}(x'^1,x'^2,w+\delta w)-{\bm r}(x^1,x^2,w)\nonumber\\
&={\bm r}(x'^1,x'^2,w+\delta w)-{\bm r'}+{\bm r'}-{\bm r},\label{kdef}\\
{\bm k'}&={\bm r}(x''^1,x''^2,w)-{\bm r}(x'^1,x'^2,w+\delta w)\nonumber\\
&={\bm r''}-{\bm r'}+{\bm r'}-{\bm r}(x'^1,x'^2,w+\delta w),\label{kpdef}
\end{align}
\end{subequations}
and according to the results (\ref{sede1})-(\ref{norwd}) the following expressions can be substituted into~(\ref{ell}) for ${\bm k}$ and ${\bm k}'$:
\begin{subequations}\label{kkkp}
\begin{align}
{\bm k}&=\delta w\, \partial_w{\bm r'}+{\bm r'}-{\bm r},\label{kkkp1}\\
{\bm k'}&=-\delta w\, \partial_w{\bm r'}+{\bm r''}-{\bm r'}.\label{kkp2}
\end{align}
\end{subequations}
Thanks to the normalization of the normal vector and its orthogonality to the tangent ones the formula (\ref{kkp}) can be rewritten in the form:
\begin{eqnarray}
&&\!\!\!\!\!\!\!\delta w\, \partial_i'{\bm r'}\!\cdot ({\bm r''}-2{\bm r'}+{\bm r})\label{eqw}\\
&&+[\partial_i'{\bm r'}\!\cdot\! ({\bm r''}-{\bm r'})]\,[\partial_w{\bm r'}\!\cdot\! ({\bm r'}-{\bm r})]\nonumber\\
&&+[\partial_i'{\bm r'}\!\cdot\! ({\bm r'}-{\bm r})]\,[\partial_w{\bm r'}\!\cdot\! ({\bm r''}-{\bm r'})]=0.\nonumber
\end{eqnarray}

In the case of thin films dealt with in the present paper all terms of the above equation can be expanded in powers of $\delta w$. From the definitions of ${\bm r}$, ${\bm r'}$ and ${\bm r''}$ (see Fig.~\ref{refle}) it stems that the quantities $\delta x^i$ and $\delta x'^i$ are of the same order too, since they are proportional to $\delta w$. In turn the difference $\delta x'^i-\delta x^i$ is right away of order of $\delta w^2$. 

For our purposes the terms up to $\delta w^4$ should be preserved in~(\ref{eqw}). The expansions are performed in the standard way. First let us concentrate on the expression
\begin{eqnarray}
{\bm r''}&&\!\!\!\!\!\!\!-2{\bm r'}+{\bm r}\simeq\delta x'^j\,\partial'_j{\bm r'} +\frac{1}{2}\,\delta x'^j\delta x'^k\, \partial'_j\partial'_k{\bm r'}\nonumber\\
&&\!\!\!\!-\delta x^j\,\partial_j{\bm r} -\frac{1}{2}\,\delta x^j\delta x^k\, \partial_j\partial_k{\bm r}.
\label{ke}
\end{eqnarray}
The third order terms are omitted since they cancel in the desired order. No higher ones are needed because of the presence of the coefficient $\delta w$ in the first expression of~(\ref{eqw}). Expanding~(\ref{ke}) further in order to get rid of the primed quantities, we obtain
\begin{eqnarray}
{\bm r''}&&\!\!\!\!\!\!\!-2{\bm r'}+{\bm r}\simeq(\delta x'^j-\delta x^j)\partial_j{\bm r}+\delta x^j\delta x^k\,\partial_j\partial_k{\bm r}\label{ke1}\\
&&\!\!\!\!+(\delta x'^j-\delta x^j)\delta x^k\,\partial_j\partial_k{\bm r}+\delta x^j\delta x^k\delta x^l\,\partial_j\partial_k\partial_l{\bm r}.
\nonumber
\end{eqnarray}
The former two terms are of order of $\delta w^2$ and the latter ones of $\delta w^3$. As to the expression $\partial_i'{\bm r'}$  standing in front of~(\ref{ke}) in~(\ref{eqw}), it is sufficient to keep the first two terms only:
\begin{equation}
\partial_i {\bm r}'\simeq \partial_i{\bm r}+\delta x^j\partial_i\partial_j{\bm r}.
\label{eksp2}
\end{equation}

Combining~(\ref{ke1}) and~(\ref{eksp2}), we can rewrite~(\ref{eqw}) in the following way
\begin{eqnarray}
&&\!\!\!\!\!\!\!\!(\delta x'^j-\delta x^j)\partial_i{\bm r}\!\cdot\!\partial_j{\bm r}+\delta x^j\delta x^k\partial_i{\bm r}\!\cdot\!\partial_j\partial_k{\bm r}\label{geqw}\\
&&= -(\delta x'^j-\delta x^j)\delta x^k\,(\partial_i{\bm r}\!\cdot\!\partial_j\partial_k{\bm r}+\partial_i\partial_k{\bm r}\!\cdot\!\partial_j{\bm r})\nonumber\\
&&-\delta x^j\delta x^k\delta x^l\,\partial_l(\partial_i{\bm r}\!\cdot\!\partial_j\partial_k{\bm r})-\frac{1}{\delta w}\,V_i(x^1,x^2,w),
\nonumber
\end{eqnarray}
where we have introduced the auxiliary vector
\begin{eqnarray}
V_i(x^1,x^2,w)=&&\!\!\!\!\!\![\partial_i'{\bm r'}\!\cdot\! ({\bm r''}-{\bm r'})][\partial_w{\bm r'}\!\cdot\! ({\bm r'}-{\bm r})]\label{vdef}\\
&&\!\!\!\!\!\!+[\partial_i'{\bm r'}\!\cdot\! ({\bm r'}-{\bm r})][\partial_w{\bm r'}\!\cdot\! ({\bm r''}-{\bm r'})]
\nonumber
\end{eqnarray}
which still has to be expanded up to $\delta w^4$. 

It is well known that the two-dimensional metric tensor is defined on the surface ${\bm r}(x^1,x^2, w)$  ($w$ being constant) as
\begin{equation}
g_{ij}=\partial_i{\bm r}\!\cdot\!\partial_j{\bm r},
\label{mete}
\end{equation}
and the Christoffel symbols of the first kind may be given the following form
\begin{equation}
\Gamma_{ijk}=\partial_i{\bm r}\!\cdot\!\partial_j\partial_k{\bm r}.
\label{chri1}
\end{equation}
If so, the formula~(\ref{geqw}) can be now rewritten as
\begin{eqnarray}
&&\!\!\!\!\!\!\!\!g_{ij}(\delta x'^j-\delta x^j)+\Gamma_{ijk}\delta x^j\delta x^k\label{geqwn}\\
&&= -(\Gamma_{ijk}+\Gamma_{jik})(\delta x'^j-\delta x^j)\delta x^k\nonumber\\
&&-\Gamma_{ijk,l}\delta x^j\delta x^k\delta x^l-\frac{1}{\delta w}\,V_i(x^1,x^2,w),
\nonumber
\end{eqnarray}
$(\ldots)_{,l}$ denoting the differentiation with respect to $x_l$. 

Now let us pass to the expression for $V_i$ and expand it up the desired order. We have to consider term by term the following factors appearing in~(\ref{vdef}). In order to save the space the expansions below are `minimal' in the sense that the terms of order $\delta w^3$ that formally should appear in~(\ref{expan1}) and~(\ref{expan2}) but cancel in the expression for $V_i$ are omitted. 
\begin{subequations}\label{expan}
\begin{align}
\partial_i'{\bm r'}&\!\cdot\! ({\bm r'}-{\bm r})\simeq\delta x^j\,\partial_i{\bm r}\!\cdot\!\partial_j{\bm r}\label{expan1}\\
&+\delta x^j\delta x^k\,\left(\frac{1}{2}\,\partial_i{\bm r}\!\cdot\!\partial_j\partial_k{\bm r}+\partial_i\partial_j{\bm r}\!\cdot\!\partial_k{\bm r}\right),\nonumber\\
\partial_i'{\bm r'}&\!\cdot\! ({\bm r''}-{\bm r'})\simeq\delta x^j\,\partial_i{\bm r}\!\cdot\!\partial_j{\bm r}\label{expan2}\\
&+\delta x^j\delta x^k\,\left(\frac{3}{2}\,\partial_i{\bm r}\!\cdot\!\partial_j\partial_k{\bm r}+\partial_i\partial_j{\bm r}\!\cdot\!\partial_k{\bm r}\right)\nonumber\\
&+(\delta x'^j-\delta x^j)\,\partial_i{\bm r}\!\cdot\!\partial_j{\bm r},\nonumber\\
\partial_w{\bm r'}&\!\cdot\! ({\bm r'}-{\bm r})\simeq-\frac{1}{2}\, \delta x^i\delta x^j\,\partial_w{\bm r}\!\cdot\!\partial_i\partial_j{\bm r}\label{expan3}\\
&+\frac{1}{2}\, \delta x^i\delta x^j\delta x^k\,\bigg(\frac{1}{3}\, \partial_w{\bm r}\!\cdot\!\partial_i\partial_j\partial_k{\bm r}\nonumber\\
&+ \partial_w\partial_i{\bm r}\!\cdot\!\partial_j\partial_k{\bm r}+ \partial_w\partial_i\partial_j{\bm r}\!\cdot\!\partial_k{\bm r}\bigg),\nonumber\\
\partial_w{\bm r'}&\!\cdot\! ({\bm r''}-{\bm r'})\simeq \frac{1}{2}\, \delta x^i\delta x^j\,\partial_w{\bm r}\!\cdot\!\partial_i\partial_j{\bm r}\label{expan4}\\
&\delta x^i(\delta x'^j-\delta x^j)\partial_w{\bm r}\!\cdot\!\partial_i\partial_j{\bm r}\nonumber\\
&+\delta x^i\delta x^j\delta x^k\,\bigg(\frac{1}{3}\, \partial_w{\bm r}\!\cdot\!\partial_i\partial_j\partial_k{\bm r}+ \frac{1}{2}\,\partial_w\partial_i{\bm r}\!\cdot\!\partial_j\partial_k{\bm r}\bigg).\nonumber
\end{align}
\end{subequations} 
Next, exploiting the transposition symmetry of the expressions like $\delta x^i\delta x^j$ as well as the property that
\begin{equation}
\partial_w{\bm r}\!\cdot\!\partial_i\partial_j{\bm r}=-\partial_w\partial_i{\bm r}\!\cdot\!\partial_j{\bm r}
\label{propp}
\end{equation}
which stems from the orthogonality of the tangent and normal vectors, one obtains
\begin{eqnarray}
&&\!\!\!\!\!\!\!\!V_i(x^1,x^2,w)=\frac{1}{2}\, g_{lk,w}\big[g_{ij}(\delta x^j\big(\delta x^k(\delta x'^l-\delta x^l)\label{vexp}\\
&&+\delta x^l(\delta x'^k-\delta x^k))-(\delta x'^j-\delta x^j)\delta x^k\delta x^l\big)\nonumber\\
&&-\Gamma_{ijn}\delta x^j\delta x^k\delta x^l\delta x^n\big].
\nonumber
\end{eqnarray}

It should be noted, that this expression is of order of $\delta w^4$, which means that the second order terms on the l.h.s of~(\ref{geqwn}) remain intact.
Collecting all terms and multiplying by the inverse metric tensor $g^{ij}$, we find the following equation for the light-ray trajectory:
\begin{eqnarray}
&&\!\!\!\!\!\!\!\!\delta x'^i-\delta x^i+\Gamma^i_{jk}\delta x^j\delta x^k\label{traje}\\
&&\!\!= -g^{il}\big[(\Gamma_{ljk}+\Gamma_{jlk})(\delta x'^j-\delta x^j)\delta x^k+\Gamma_{ljk,m}\delta x^j\delta x^k\delta x^m\big]\nonumber\\
&&\!\!-\frac{1}{2\delta w}\,g_{lk,w}\big[\delta x^i\big(\delta x^k(\delta x'^l-\delta x^l)+\delta x^l(\delta x'^k-\delta x^k))\nonumber\\
&&\!\!-(\delta x'^i-\delta x^i)\delta x^k\delta x^l-\Gamma^i_{jm}\delta x^j\delta x^k\delta x^l\delta x^m\big],\nonumber
\end{eqnarray}
Up to the considered order one can replace the expressions $\delta x'^i-\delta x^i$ with
\begin{equation}
\delta x'^i-\delta x^i\longmapsto -\Gamma^i_{jk}\delta x^j\delta x^k,
\label{rep}
\end{equation}
which leads to the final formula
\begin{eqnarray}
&&\!\!\!\!\!\!\!\!\delta x'^i-\delta x^i+\Gamma^i_{jk}\delta x^j\delta x^k\label{traje2}\\
&&= g^{il}\big[(\Gamma_{lnj}+\Gamma_{nlj})\Gamma^n_{km}-\Gamma_{ljk,m}\big]\delta x^j\delta x^k\delta x^m\nonumber\\
&&+\delta w^{-1}g_{lk,w}\Gamma^l_{jm}\delta x^i\delta x^j\delta x^k\delta x^m,\nonumber
\end{eqnarray}
where $\Gamma^i_{jk}=g^{il}\Gamma_{ljk}$ are Christoffel symbols of the second kind.

In order to obtain the differential equations for coordinates $x^1, x^2$ on the surface, one has to observe that $\delta w$ plays the double role in our considerations. Firstly it is connected with the thickness of the considered layer and secondly it defines the `infinitesimal' step while passing from $x^i$ to $x^i+\delta x^i$. Therefore whenever the quantity $\delta x^i/\delta w$ appears it can be replaced with $d x^i/dt$, where for stressing this special role the name of the parameter along the curve has been changed into $t$. In turn the quantity $(\delta x'^j-x^j)/(\delta w)^2$ is just the second symmetric derivative, which for twofold differentiable function simply equals $d^2x^i/dt^2$~\cite{secsym}. In that way one gets the differential equation for the light-ray trajectory:
\begin{eqnarray}
&&\!\!\!\!\!\!\!\!\frac{d^2 x^i}{dt^2}+\Gamma^i_{jk}\frac{d x^j}{dt}\frac{d x^k}{dt}= \delta w\bigg\{g^{il}\bigg[(\Gamma_{lnj}+\Gamma_{nlj})\Gamma^n_{km}\label{trajer}\\
&&-\Gamma_{ljk,m}\frac{d x^j}{dt}\frac{d x^k}{dt}\frac{d x^m}{dt}\bigg]+g_{lk,w}\Gamma^l_{jm}\frac{d x^i}{dt}\frac{d x^j}{dt}\frac{d x^k}{dt}\frac{d x^m}{dt}\bigg\}.\nonumber
\end{eqnarray}

For very thin layers, i.e. for $\delta w\rightarrow 0$, the right-hand side disappears, and the equation of a geodesic is obtained, as it might be expected:
\begin{equation}
\frac{d^2 x^i}{dt^2}+\Gamma^i_{jk}\frac{dx^j}{dt}\,\frac{dx^k}{dt}=0.
\label{eqgeod}
\end{equation}
For more thick layers the expression on the r.h.s. of~(\ref{trajer}) constitutes the correction to the geodesic and modifies the trajectory.
The first term, that contains the third power of derivatives with respect to the parameter $t$ comes from the distinct inclination of the normal vector $\partial_w{\bm r}$ at different points of reflection and the term with fourth power of derivatives is the consequence of the different length of vectors $\bm k$ and $\bm k'$.

The terms standing on the r.h.s. o~(\ref{trajer}) as compared to $\Gamma^i_{jk}\dot{x}^j\dot{x}^k$ can be estimated to be of order of $\delta g/g$ and $\delta g'/g'$, where $\delta g$ ($\delta g'$) denotes the change of a given element of the metric tensor (or its derivative) while moving tangentially along the surface by the distance corresponding to $2\delta w$ (strictly speaking it is the distance between the points $(x^1,x^2,w)$ and $(x''^1,x''^2,w)$ in Fig.~\ref{refle}). This estimate holds for truly curved surfaces, where all Christoffel symbols cannot simultaneously vanish. For surfaces without intrinsic curvature as for instance the cylindrical one only the second-derivative term survives leading to the trivial equation. It should be emphasized that within geometrical optics there is no length scale imposed by the wavelength of the propagating light. The only such scales refer to the layer thickness and its normal curvatures so the appearance of the quantities such as $\delta g/g$ might have been expected.

\begin{figure*}[t]
\begin{center}
\includegraphics[width=0.98\textwidth,angle=0]{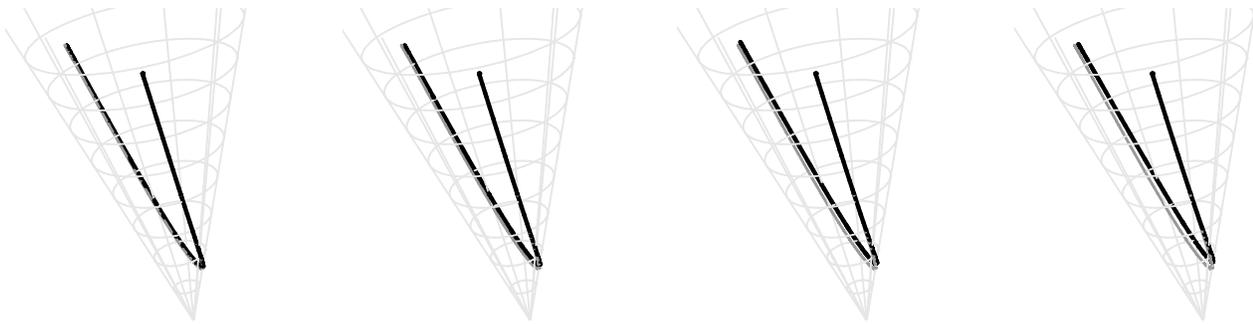}
\end{center}
\caption{The light-ray trajectories (black lines) generated out of~(\ref{eqmcone}) as compared to the geodesic (gray line) for a thin cone layer. The exemplary cone surface corresponds to the value $w=0$. The subsequent values of $\delta w$ are $0.01,\, 0.02,\, 0.03,\, 0.04$ for which $\delta w/\rho=0.005,0.01,0.015,\, 0.02$ if calculated at the starting point of each trajectory. When approaching to the cone tip these values increase proportionally to $1/\rho$.}
\label{con}
\end{figure*}

\section{Examples}
\label{num}

\subsection{Cone}
\label{cone}

An exemplary cone surface can be described by the relation ${\bm r}(\rho,\phi, w)$ ($w$ being constant) with 
\begin{subequations}\label{conec}
\begin{align}
x(\rho,\phi,w)&=(\rho+w/\sqrt{2})\cos\phi,\label{xcone}\\
y(\rho,\phi,w)&=(\rho+w/\sqrt{2})\sin\phi,\label{ycone}\\
z(\rho,\phi,w)&=\rho-w/\sqrt{2}.\label{zcone}
\end{align}
\end{subequations}
The requirements~(\ref{norm}) and~(\ref{ortog}) are satisfied in the obvious way. On the conical surface the parameter $\rho$ plays the role of the coordinate $x^1$ and $\phi$ that of $x^2$. It can be easily verified that the cone equation is satisfied:
\begin{equation}
x^2+y^2=(z+w\sqrt{2})^2.
\label{coneq}
\end{equation}
with the apex at $z=-\sqrt{2}\,w$. In these coordinates the metric tensor has the form
\begin{equation}
g_{ij}=\left[\begin{array}{cc}2 & 0 \\ 0 & (\rho + w/\sqrt{2})^2\end{array}\right],
\label{metr}
\end{equation}
and the only nonzero derivatives of its elements are
\begin{equation}
g_{\phi\phi,\rho}=2(\rho + w/\sqrt{2}),\;\;\;\; g_{\phi\phi,w}=\sqrt{2}\,\rho + w.
\label{metrdif}
\end{equation}
Now the Christoffel symbols can be easily calculated. The only non-vanishing ones of the first kind are
\begin{equation}
\Gamma_{\phi\phi\rho}=\Gamma_{\phi\rho\phi} =\rho + w/\sqrt{2},\;\;\; \Gamma_{\rho\phi\phi} =-(\rho + w/\sqrt{2}). \label{cr1}
\end{equation}
and of the second kind
\begin{equation}
\Gamma^\phi_{\phi\rho}=\Gamma^\phi_{\rho\phi} =(\rho + w/\sqrt{2})^{-1},\;\;\; \Gamma^\rho_{\phi\phi} =-\frac{1}{2}(\rho + w/\sqrt{2}).\label{cr2}
\end{equation}
In order to explicitly write out~(\ref{trajer}) we will also need the values of certain derivatives:
\begin{equation}
\Gamma_{\phi\phi\rho,\rho}=\Gamma_{\phi\rho\phi,\rho} =1,\;\;\;\; \Gamma_{\rho\phi\phi,\rho} =-1.
\label{dega}
\end{equation}
This leads to the following trajectory equations:
\begin{subequations}\label{eqmcone}
\begin{align}
\frac{d^2 \rho}{dt^2}-&\frac{1}{2}(\rho + w/\sqrt{2})\left(\frac{d\phi}{dt}\right)^2=\label{eqmcone1}\\
&\delta w \bigg[\frac{1}{2}\frac{d\rho}{dt}\left(\frac{d\phi}{dt}\right)^2+2\sqrt{2} \left(\frac{d\rho}{dt}\right)^2\left(\frac{d\phi}{dt}\right)^2\bigg],\nonumber\\
\frac{d^2 \phi}{dt^2}+&\frac{2}{\rho + w/\sqrt{2}}\frac{d\rho}{dt}\!\cdot\!\frac{d\phi}{dt}=\label{eqmcone2}\\
&\delta w \bigg[\frac{2}{(\rho + w/\sqrt{2})^2}\left(\frac{d\rho}{dt}\right)^2\frac{d\phi}{dt}+2\sqrt{2}\, \frac{d\rho}{dt}\left(\frac{d\phi}{dt}\right)^3\bigg].\nonumber
\end{align}
\end{subequations}

In Fig.~\ref{con} the light-ray trajectories (plotted in black), which represent the solutions of the full equations~(\ref{eqmcone}), are provided for increasing values of $\delta w$. It can be observed that for thin layer the light ray follows the geodesic (plotted in gray). For larger values of the layer thickness the true trajectory slightly deviates from the geodesic. Especially it happens close to the cone apex, where one of the normal curvatures becomes large.
It is understandable since the right-hand sides of~(\ref{eqmcone}), which constitute corrections to the geodesic equation, turn out to be of order $\delta w/\rho$ (for a path perpendicular to the symmetry axis) as compared to the second terms on the left-hand sides, i.e. those in~(\ref{eqgeod}) containing Christoffel symbols. Therefore, one can conclude that, roughly speaking, whenever one thinks about small value of $\delta w$ that means `small as compared to the normal curvature radii'. This is confirmed by the presented plots  because the deviation from the geodesic manifests mainly for small values of $\rho$.

For larger values of $\delta w$ the trajectories become somewhat chaotic: the small modifications of the parameters of the incoming ray (the direction of the ray or its inclination with respect to the surface) result in significantly distinct trajectories. This is the typical behavior for nonlinear differential equations like~(\ref{trajer}). The same observations refer to the subsequent plots involving other layers.

\begin{figure*}[t]
\begin{center}
\includegraphics[width=0.98\textwidth,angle=0]{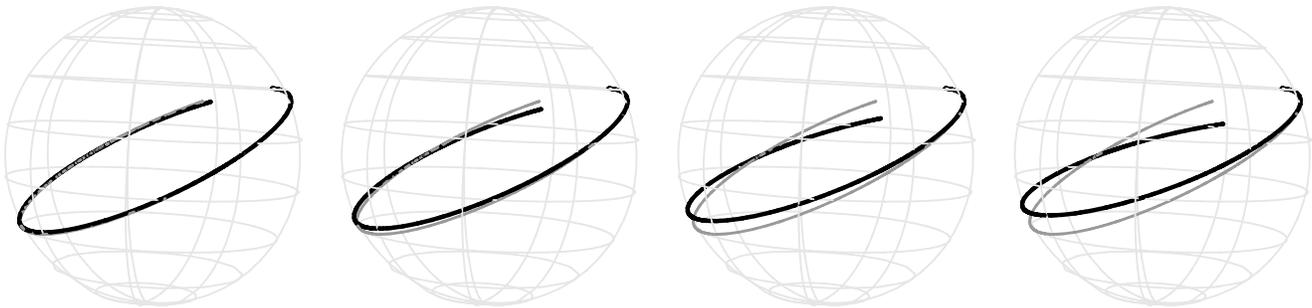}
\end{center}
\caption{Same as Fig.~\ref{con} but for a thin spherical layer. For the limiting surface $R=1$, $w=0$ is chosen. The subsequent values of $\delta w$ are $0.01,\, 0.03,\, 0.07,\, 0.1$ which are simultaneously equal to $\delta w/R$.}
\label{sph}
\end{figure*}

\subsection{Sphere}
\label{sphere}

The parametrization of the sphere satisfying~(\ref{norm}) and~(\ref{ortog}) is easy to write down:
\begin{subequations}\label{sphd}
\begin{align}
&x(\theta,\phi,w)=(R+w)\sin\theta\cos\phi,\label{xsph}\\
&y(\theta,\phi,w)=(R+w)\sin\theta\sin\phi,\label{ysph}\\
&z(\theta,\phi,w)=(R+w)\cos\theta,\label{zsph}
\end{align}
\end{subequations}
where $x^1=\theta$, $x^2=\phi$ and $R$ is the sphere radius. The metric tensor on the sphere has the standard form, apart from the presence of $w$:
\begin{equation}
g_{ij}=\left[\begin{array}{cc}(R + w)^2& 0\\ 0& (R + w)^2 \sin^2\theta\end{array}\right],
\label{metrsph}
\end{equation}
and performing the simple differentiation one gets
\begin{eqnarray}
g_{\theta\theta,w}&&\!\!\!\!=2(R + w),\;\;\;\; g_{\phi\phi,w}=2(R + w) \sin^2\theta,\nonumber\\
g_{\phi\phi,\theta}&&\!\!\!\!=(R + w)^2 \sin 2\theta. \label{metrsphd}
\end{eqnarray}

From that the known expressions for the Christoffel symbols can be obtained:
\begin{equation}
\Gamma_{\phi\phi\theta}=\Gamma_{\phi\theta\phi}=-\Gamma_{\theta\phi\phi}=\frac{1}{2}\,(R + w)^2 \sin 2\theta,
\label{chsp1}
\end{equation}
and also
\begin{equation}
\Gamma^\phi_{\phi\theta}=\Gamma^\phi_{\theta\phi}=\cot\theta,\;\;\;\; \Gamma^\theta_{\phi\phi}=-\frac{1}{2}\,\sin 2\theta,
\label{chsp2}
\end{equation}
the others being zero. The only derivatives that come into play, are those over $\theta$:
\begin{equation}
\Gamma_{\phi\phi\theta,\theta}=\Gamma_{\phi\theta\phi,\theta}=-\Gamma_{\theta\phi\phi,\theta}=(R + w)^2 \cos 2\theta.
\label{chspd}
\end{equation}
The light-ray trajectory equations~(\ref{trajer}) take the following form
\begin{subequations}\label{eqsph}
\begin{align}
\frac{d^2 \theta}{dt^2}&-\frac{1}{2}\,\sin 2\theta\left(\frac{d\phi}{dt}\right)^2=\delta w\bigg[\cos 2\theta\,\frac{d\theta}{dt} \left(\frac{d\phi}{dt}\right)^2\nonumber\\
&+(R+w)\sin 2\theta\left(\frac{d\theta}{dt}\right)^2 \left(\frac{d\phi}{dt}\right)^2\bigg],\label{eqsph1}\\
\frac{d^2 \phi}{dt^2}&+2\cot\theta\, \frac{d\theta}{dt} \cdot\frac{d\phi}{dt}=\delta w\bigg[\frac{2}{\sin^2\theta} \left(\frac{d\theta}{dt}\right)^2\frac{d\phi}{dt}\nonumber\\
&+(R+w)\sin 2\theta\,\frac{d\theta}{dt} \left(\frac{d\phi}{dt}\right)^3\bigg].\label{eqsph2}
\end{align}
\end{subequations}

It can be shown that the r.h.s is smaller with respect to the l.h.s by the factor $\delta \theta$ (or similarly $\delta\phi$): the change in the angle $\theta$ after two reflections of the light ray, as shown in Fig.~\ref{refle}. Naturally it is the quotient of the arc length and the radius, which obviously is of the order of $\delta w/R$. Once again we come to the conclusion that the thickness of the layer should be referred to the normal curvature radius.

The deviations of the trajectory from the geodesic, which is a great circle, are shown in Fig.~\ref{sph} for subsequently increasing values of the layer thickness. The incoming ray is chosen as oriented along a `parallel'. As in the case of the cone the deviations from a geodesic increase with growing $\delta w$. Due to the symmetry no deviations would be observed if the incident ray were moving along a great circle.

\subsection{Torus}
\label{torus}

\begin{figure*}[t!]
\begin{center}
\includegraphics[width=0.98\textwidth,angle=0]{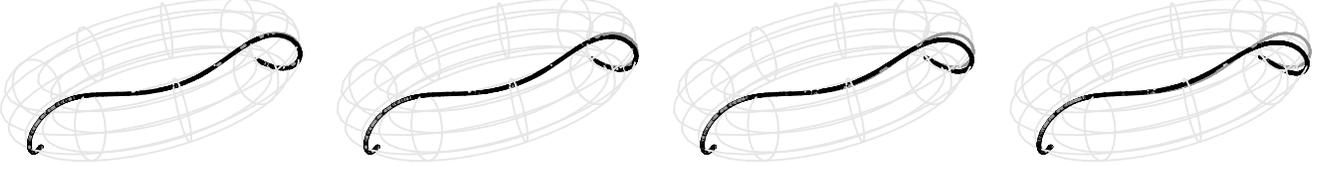}
\end{center}
\caption{Same as Fig.~\ref{con} but for a thin toroidal layer. For the limiting torus surface $w=2$ and $R=10$. The subsequent values of $\delta w$ are $0.005,\, 0.01,\, 0.02,\, 0.03$.}
\label{tor}
\end{figure*}

The coordinates describing an exemplary torus may be chosen to be:
\begin{subequations}\label{toru}
\begin{align}
&x(\theta,\phi,w)=(R+w \sin\theta)\cos \phi,\label{xtor}\\
&y(\theta,\phi,w)=(R+w \sin\theta)\sin \phi,\label{ytor}\\
&z(\theta,\phi,w)=w \cos\theta.\label{ztor}
\end{align}
\end{subequations}
where $R$ is a parameter. It is straightforward to verify that the conditions~(\ref{norm}) and~(\ref{ortog}) are fulfilled. As before the angle $\theta$ corresponds to $x^1$ and $\phi$ to $x^2$.

The metric tensor on this surface can be easily calculated:
\begin{equation}
g_{ij}=\left[\begin{array}{cc}w^2 & 0 \\ 0 & (R + w\sin\theta)^2\end{array}\right].
\label{metrt}
\end{equation}
and the only nonzero derivatives of its elements are
\begin{equation}
g_{\phi\phi,\theta}=2w(R + w\sin\theta)\cos\theta,\;\;\;\; g_{\phi\phi,w}=2w.
\label{metrdift}
\end{equation}
For the Christoffel symbols of the first kind the following expressions are obtained:
\begin{eqnarray}
&&\Gamma_{\phi\phi\theta}=\Gamma_{\phi\theta\phi} =w(R+w\sin\theta)\cos\theta,\nonumber\\
&&\Gamma_{\theta\phi\phi} =-w(R+w\sin\theta)\cos\theta, \label{cr1t}
\end{eqnarray}
with the derivatives
\begin{equation}
\Gamma_{\phi\phi\theta,\theta}=\Gamma_{\phi\theta\phi,\theta}=-\Gamma_{\theta\phi\phi,\theta}=w^2\cos 2\theta-Rw\sin\theta.
\label{gvc}
\end{equation}
The second kind symbols will also be needed:
\begin{eqnarray}
&& \Gamma^\phi_{\phi\theta}=\Gamma^\phi_{\theta\phi} =\frac{w\cos\theta}{R+w\sin\theta},\nonumber\\
&& \Gamma^\theta_{\phi\phi} =-\frac{(R+w\sin\theta)\cos\theta}{w}.\label{crt2}
\end{eqnarray}

Now we are in a position to assemble our fundamental differential equations~(\ref{trajer}):
\begin{subequations}\label{eqmt}
\begin{align}
\frac{d^2 \theta}{dt^2}-&\frac{(R+w\sin\theta)\cos\theta}{w}\left(\frac{d\phi}{dt}\right)^2=\label{eqmt1}\\
&\delta w \bigg[\left(\cos 2\theta -\frac{R\sin\theta}{w}\right)\frac{d\theta}{dt}\left(\frac{d\phi}{dt}\right)^2\nonumber\\
&+2\cos\theta(w\sin\theta-R)\left(\frac{d\theta}{dt}\right)^2\left(\frac{d\phi}{dt}\right)^2\bigg],\nonumber\\
\frac{d^2 \phi}{dt^2}+&\frac{2w\cos\theta}{R+w\sin\theta}\frac{d\theta}{dt}\!\cdot\! \frac{d\phi}{dt}=\label{eqmt2}\\
&\delta w \bigg[ 2w\,\frac{w+R\sin\theta}{(R+w\sin\theta)^2}\left(\frac{d\theta}{dt}\right)^2\frac{d\phi}{dt}\nonumber\\
&+2\cos\theta(w\sin\theta-R)\frac{d\theta}{dt}\,\left(\frac{d\phi}{dt}\right)^3\bigg].\nonumber
\end{align}
\end{subequations}

It can be checked that in order to  be allowed to neglect the right-hand sides, the quantities $\delta w/R$ and $\delta w/w$ have to be tiny, which means that the principal curvature radii should be large with respect to $\delta w$. The resulting trajectories for various values of $\delta w$ are shown in Fig.~\ref{tor}. They confirm the observations made for the cone and sphere: the thicker the layer the more the curves depart from the geodesic. When changing the initial parameters for the light ray their chaotic character can also be revealed. If the incident ray is oriented along the curve corresponding to $\theta=\pm \pi/2$ or $\phi=\mathrm{const}$ no deviations are observed.

\subsection{Catenoid}
\label{catenoid}

\begin{figure*}[t!]
\begin{center}
\includegraphics[width=0.98\textwidth,angle=0]{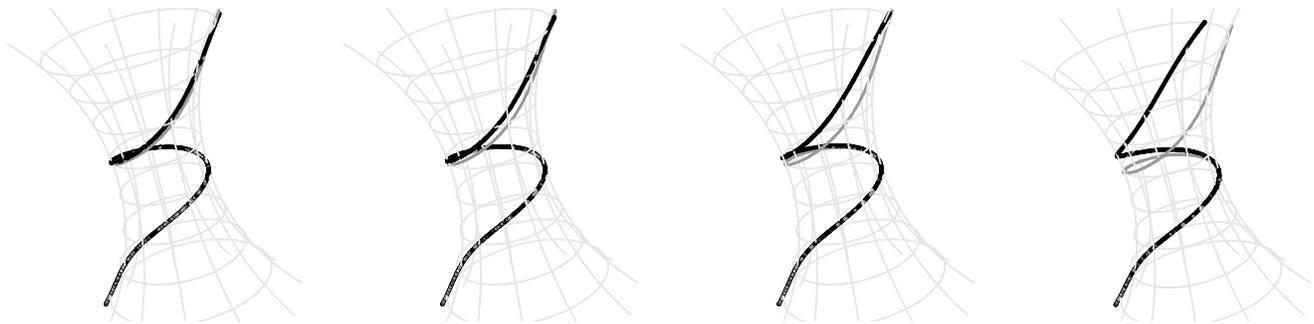}
\end{center}
\caption{Same as Fig.~\ref{con} but for a thin catenoid layer. For the limiting catenoid surface $w=1$, and the scale factor has been chosen to be $a=2$. The subsequent values of $\delta w$ are $0.0005,\, 0.001,\, 0.002,\, 0.003$ which simultaneously equal to $2\delta w/a$.}
\label{cat}
\end{figure*} 

An interesting example constitutes the catenoid which is a minimal surface. The appropriate orthogonal coordinates describing an exemplary catenoid and satisfying the normalization of the normal vector can be found to be~\cite{barg}:
\begin{subequations}\label{cate}
\begin{align}
&x(u,v,w)=\left(a\cosh u-\frac{w}{\cosh u}\right)\cos v,\label{xcat}\\
&y(u,v,w)=\left(a\cosh u-\frac{w}{\cosh u}\right)\sin v,\label{ycat}\\
&z(u,v,w)=a u +w\tanh u.\label{zcat}
\end{align}
\end{subequations}
where $a$ is a parameter for our choice and $u=x^1$ and $v=x^2$.

By calculation of the products of the tangent vectors, the metric tensor can easily be determined:
\begin{equation}
g_{ij}=\left[\begin{array}{cc} \displaystyle \left(a\cosh u+\frac{w}{\cosh u}\right)^2 & 0\\ 0 &\displaystyle \left(a\cosh u-\frac{w}{\cosh u}\right)^2\end{array}\right].
\label{mecat}
\end{equation}
Its elements depend only on $u$ and $w$ and the derivatives with respect to these parameters are
\begin{eqnarray}
g_{uu,u}&&\!\!\!\!=g_{vv,u}=2\sinh u\left(a^2\cosh u-\frac{w^2}{\cosh^3 u}\right),\nonumber\\
g_{uu,w}&&\!\!\!\!=2\left(a+\frac{w}{\cosh^2 u}\right),\nonumber\\
g_{vv,w}&&\!\!\!\!=2\left(-a+\frac{w}{\cosh^2 u}\right).
\label{pogu}
\end{eqnarray}
They allow to write down the non-vanishing Christoffel symbols
\begin{eqnarray}
\Gamma_{uuu}&&\!\!\!\!=\Gamma_{vuv}=\Gamma_{vvu}=-\Gamma_{uvv}\label{chsc1}\\
&&\!\!\!\!=\left(a^2\cosh u-\frac{w^2}{\cosh^3 u}\right)\sinh u,\nonumber
\end{eqnarray}
and
\begin{eqnarray}
\Gamma^u_{uu}&&\!\!\!\!=-\Gamma^u_{vv}=\frac{2a \sinh 2u}{a(1+\cosh 2u)+2w}-\tanh u,\label{chsc2}\\
\Gamma^v_{uv}&&\!\!\!\!=\Gamma^v_{vu}=\frac{a \sinh 2u+2w \tanh u}{a(1+\cosh 2u)-2w}.\nonumber
\end{eqnarray}
For the derivatives of~(\ref{chsc1}) we get
\begin{eqnarray}
\Gamma_{uuu,u}&&\!\!\!\!=\Gamma_{vuv,u}=\Gamma_{vvu,u}=-\Gamma_{uvv,u}\label{chscde}\\
&&\!\!\!\!=a^2 \cosh 2u + w^2\,\frac{\cosh 2u-2}{\cosh^4u}.\nonumber
\end{eqnarray}

In order to obtain the explicit equations for the trajectory the above quantities should be now plugged into~(\ref{trajer}). We are not going, however, to write down explicitly these lengthy expressions and limit ourselves to the presentation of the corresponding curves. They are depicted in Fig.~\ref{cat}. Here the depart of the black curve from the geodesic manifests itself more strongly. The visible deviation starts when the light ray enters the region of maximal narrowing and therefore also maximizing one of the principal curvatures (measured there by $1/a$).

\section{Summary}

In the present work our concern were the light-ray trajectories in thin layers. It was assumed that these rays are reflected by the boundaries (for instance due to the phenomenon of the total internal reflection in a dielectric layer or by reflecting walls) which in general are surfaces of nontrivial curvature. At each point where the ray hits the boundary the usual law of reflection is satisfied: the incident ray, the reflected ray and the normal vector lie in the same plane and the reflection angle equals the incidence angle. This law is sufficient to reconstruct the whole trajectory of the ray.

The differential equation of the trajectory (\ref{trajer}) has been derived. This is a highly nonlinear equation with the thickness (measured by $\delta w$) as a parameter. It has been shown that when setting $\delta w\longrightarrow 0$, which corresponds to infinitely thin layer, the trajectory equation reduces to that of a geodesic drawn on the boundary surface (in this case both boundaries merge into one). 

When $\delta w$ increases and the layer has certain non-negligible thickness the corrections (i.e. the r.h.s. of (\ref{trajer})) start playing a role leading to the depart of the traveling ray from the geodesic. This effect becomes stronger as the layer thickness increases. 

The special cases dealt with in section \ref{num} confirm these observations. The plots performed for four truly curved layers show the observable deviation of the trajectory from the geodesic. This effect is magnified if the normal curvature connected with the appropriate path becomes large. The essential parameter seems to be $\delta w/R$, where $R$ denotes the normal curvature radius at a given point and for a given curve. However, the expression standing on the r.h.s. of~(\ref{trajer}) and involving Christoffel symbols cannot be in a simple way expressed by the principal curvatures. The dependence of the ray trajectory on curvatures, although obvious, has rather complicated nature. The findings of our works stay in general agreement with experimental results~\cite{schul}. 

The outcome of this work is expected to be confirmed by calculating of the Poynting vector for the wave beam propagating in a layer, which is, however, mathematically nontrivial particularly for complicated surfaces. The role of the wavelength (absent in geometrical optics) would then be clarified. Our present findings contribute to the discussion~\cite{vankamp,silva,brb} to what extent thin layers may be treated as two-dimensional structures (thereby, pointing some additional restrictions),  constitute the step toward better understanding how to manipulate the light in thin films and try to explain the role of curvature. 

\section*{Acknowledgments}
I would like to thank to Professor Iwo Bia\l ynicki-Birula for the inspiration and elucidating discussions.


\begin{thebibliography}{99}
\bibitem{tam1} T. Tamir (ed.), {\em Integrated Optics}, Springer, New York, 1979.
\bibitem{hun} R.G. Hunsperger, {\em Integrated Optics: Theory and Technology}, Springer, Berlin 1985.
\bibitem{tam2} T. Tamir (ed.), {\em Guided- Wave Optoelectronics}, Springer, New York 1990.
\bibitem{sc} R.R.A. Syms and J.R. Cozens, {\em Optical Guided Waves and Devices}, McGraw-Hill, 1992.
\bibitem{yeh} C. Yeh and F. Shimabukuro, {\em The Essence of Dielectric Waveguides}, Springer, New York 2008.
\bibitem{bal} C.A. Balanis, {\em Advanced Engineering Electromagnetics}, Wiley, New York 1989.
\bibitem{carson} J.R. Carson, S.P. Mead and S.A. Schelkunoff, Bell Syst. Tech. J. {\bf 15}, 310(1936).
\bibitem{chew} W.C. Chew, {\em Waves and Fields in Inhomogeneous Media}, Von Nostrand Reinhold, New York 1990. 
\bibitem{collin} R.E. Collin, {\em Field Theory of Guided Waves}, IEEE Press, New York 1990.
\bibitem{marcuse} D. Marcuze, {\em Theory of dielectric optical waveguides}, Academic
Press, New York 1974.
\bibitem{hillion} P. Hillion, Pure Appl. Opt. {\bf 1}, 169(1992).
\bibitem{lapidus} I.R. Lapidus,  Am. J. Phys {\bf 50}, 155(1982).
\bibitem{zwiebach} B. Zwiebach, {\em A First Course in String Theory}, Cambridge University Press, Cambridge 2009.
\bibitem{trq}  T. Rado\.zycki, arXiv:1802.02062 (2018).
\bibitem{brb} T. Rado\.zycki and P. Bargie{\l}a, J. Mod. Opt. {\bf 65}, 1404((2018).
\bibitem{batz} S. Batz and U. Peschel, Phys. Rev. {\bf A 78}, 043821(2008).
\bibitem{lai} M.-Y. Lai, Y.-L. Wang, G.-H. Liang, F. Wang and H,-S. Zong, Phys. Rev. {\bf A97}, 033843(2018)
\bibitem{bra} J.C. Bradley, E.C. Malarkey, D. Mergerian, H.A. Trenchard, {\em Theory Of Geodesic Lenses}, Proc. SPIE 0176, {\em Guided Wave Optical Systems and Devices II}, (26 July 1979).
\bibitem{li} L. Xu, X. Wang, T. Tyc, Ch. Sheng, S. Zhu, H. Liu and H. Chen, arXiv:1801.10438 (2018).
\bibitem{ri} G.C. Righini, V. Russo, S. Sottini, and G. Toraldo di Francia, Applied Optics {\bf 12}, 1477(1973).
\bibitem{corn} S. Cornbleet, P. Rinous, {\em Generalised formulas for equivalent geodesic and nonuniform
refractive lenses}. IEE Proceedings H-Microwaves, Optics and Antennas, 1981, IET 1981.
\bibitem{sar}  M. \v{S}arbort and T. Tyc, J. Opt.{\bf 14}, 075705(2012).
\bibitem{schul} V.H. Schultheiss, S. Batz, A. Szameit, F. Dreisow, S. Nolte,A.  T\"unnermann,S. Longhi, and U. Peschel, Phys. Rev. Lett. {\bf 105}, 143901(2010).
\bibitem{secsym} B.S. Thomson, {\em Symmetric properties of real functions}, Marcel Dekker Inc., New York 1994.
\bibitem{barg} P. Bargie{\l}a, private communication.
\bibitem{vankamp} N.G. van Kampen and J.J. Lodder, Am. J. Phys. {\bf 52}, 419(1984).
\bibitem{silva} L.C.B. da Silva, C.C. Bastos, F.G.Ribeiro, Ann. Phys. (New York) {\bf 379}, 13(2017).
\end{thebibliography}
\end{document}